\documentclass{article}

\usepackage{PRIMEarxiv}

\usepackage[utf8]{inputenc} 
\usepackage[T1]{fontenc}    
\usepackage{hyperref}       
\usepackage{url}            
\usepackage{booktabs}       
\usepackage{amsfonts}       
\usepackage{nicefrac}       
\usepackage{microtype}      
\usepackage{lipsum}
\usepackage{graphicx}
\usepackage{amsmath}
\usepackage{amssymb,amsthm,amsmath}
\usepackage{xcolor,paralist,hyperref,titlesec,fancyhdr,etoolbox}
\graphicspath{{media/}}     

\title{Cryptanalysis of Nikooghadam et al.'s lightweight Authentication Protocol for Internet of Drones
 
}

\author{
  Iman Jafarian \\
  Department of Computer Engineering, Amirkabir University of Technology, Tehran, Iran \\
  \texttt{iman.j@aut.ac.ir } \\
}

\begin{document}
\maketitle

\begin{abstract}
The Internet of Drones has emerged as a transformative technology with applications spanning various domains, including surveillance, delivery services, and disaster management. Secure communication between controller users and drones is paramount to ensure the transmitted data's confidentiality, integrity, and authenticity. Key agreement protocols are crucial in establishing secure communication channels between users and drones, enabling them to exchange sensitive information and control their operations securely. Recently Nikooghadam et al. proposed a lightweight mutual authentication and key agreement protocol for the Internet of drones. In this article, we provide a descriptive analysis of their proposed scheme and prove that Nikooghadam et al.'s scheme is vulnerable to user tracking attacks and stolen verifier attacks.
\end{abstract}


\section{Introduction}
The Internet of Things (IoT) has revolutionized how we interact with technology and our surroundings, connecting various devices and enabling seamless communication and data exchange. Within the realm of IoT, the Internet of Drones (IoD) has emerged as an innovative and transformative concept, integrating unmanned aerial vehicles (UAVs) or drones into the interconnected network. The Internet of Drones opens up many possibilities and applications across different sectors, including surveillance, delivery services, agriculture, infrastructure inspection, disaster management, and more. Drones equipped with advanced sensors, cameras, and communication capabilities can gather real-time data, perform complex tasks, and operate in challenging environments where human intervention is limited.
Key agreement protocols in the IoD leverage cryptographic techniques to establish secure communication channels. Public key cryptography, specifically elliptic curve cryptography (ECC), is commonly used due to its strong security properties and suitability for resource-constrained devices like drones, and many of the last published papers have used this cryptographic technique.

Despite all the benefits of the Internet of Drones (IoD), establishing a secure channel for communication is a major challenge. Failure to establish a secure channel in this environment can lead to various negative consequences, including compromised data security, unauthorized drone control, privacy breach, and trust and reputation damage. Hence, research has recently focused on providing IoD secure authentication and communication schemes. [1, 2, 3, 4, 5, 6]

\section{Review and cryptanalysis of Nikooghadam et al.'s scheme}
\label{sec:headings}

In this section, we review and analyze Nikooghadam et al.’s scheme [6], demonstrating that it suffers from user tracking and stolen verifier attacks.
\subsection{review of Nikooghadam et al.'s scheme}
The notations used in the scheme are shown in Table 1. This scheme contains two main phases: registration and authentication. In the registration phase, the controller user and the drone register in the control server. User and drone registration phases are shown in Figure 1 and Figure 2, respectively. Then the controller user shares a key with the drone via the controller server in the authentication phase—the shared key is used for their subsequent secure communications. The steps of this phase are shown in Figure 3.

\begin{table}
	\scriptsize
	\centering
	\caption{Notations used Nikooghadam et al.' scheme [6]}\label{tab1}
	\begin{tabular}{|c|c|}
		\hline
		\textbf{Notaion} & \textbf{Description} \\
		\hline 
		$U_i$ & $i$-th User  \\ 
		$V_j$ & $j$-th Drone \\ 
		$ID_i$ & Identity of $U_i$  \\
		$ID_j$ & Identity of $V_j$  \\
		$CS$ & Control Server \\
		$P$ & Base point of $E_p(a, b)$ \\
		$s$ & The secret key of $CS, s \in Z_p$  \\
		$sk$ & Session keys  \\ 
		$T_x$ & Timestamp (1 $\leq x \leq $ 4) \\
		$\Delta T$ & Threshold value for the timestamp (1 $\leq x \leq $ 4) \\
		$h(.)$ & Hash function \\
		$\oplus$ & XOR operation  \\
		$||$ & Concatenation operator  \\
		$a_j, d_i, q_i, z_i, g_j$ & Numbers selected from $Z_p$  \\
		\hline 
	\end{tabular} 
\end{table}

\begin{figure}[!htbp]
\centering \scalebox{0.9}{
\begin{tabular}{|l l l|}
\hline
~~~~~~~~~~~~\textbf{User $U_i$/Mobile device}      & \textbf{SecureChannel~~~}    & ~~~~~~~~~~~~~\textbf{~~~~Control server($CS$)}  \\
\hline
Select identity $ID_i$ and password $PW_i$    & &  \\
Select random number $d_i\in Z_p$ & & \\
Compute $ppw_i = h(h(ID_i||d_i)\oplus h(PW_i||d_i))$ & & \\
&          $~~\underrightarrow{\{ID_i, ppw_i\}}$       &  \\
& & Select two random numbers $f_i, q_i \in Z_p$ \\
& & Compute $FID_i=h(ID_i||f_i)$           \\
& & Compute $K_i=h(FID_i||s||q_i)$  \\
& & Compute $A_i=h(FID_i||ppw_i||f_i||K_i)$ \\
& & Compute $B_i=h(A_i||FID_i)$ \\
& & Store ($ID_i, FID_i, K_i$) in the database      \\
&         $\underleftarrow {\{f_i, K_i, B_i, h(.)\}}$       &  \\
Store $\{d_i, f_i, K_i, B_i, h(.)\}$ in the mobile device & & \\
\hline
\end{tabular}}
\caption{user registration of Nikooghadam et al.' scheme [6]} \label{fig:UPReg}
\end{figure}

\begin{figure}[!htbp]
\centering \scalebox{0.9}{
\begin{tabular}{|l l l|}
\hline
~~~~~~~~~~~~~~~\textbf{Drone $V_j$}      &  \textbf{SecureChannel}   & ~~~~~~~~~~~~~~~\textbf{Control server($CS$)}  \\
\hline
Select identity $ID_j$    & &  \\

&         $~~~~\underrightarrow{~~~~~\{ID_j\}~~~~~}$       &  \\
& & If $ID_j$ is in database \\
& & ~~~~~~~ Request another unique identity \\
& & Else           \\
& & ~~~~~~~Select random number $a_j \in Z_p$  \\
& & ~~~~~~~Compute $PID_j=h(a_j||ID_j)$ \\
& & ~~~~~~~Compute $Key_j=h(ID_j||s||a_j)$ \\
& & ~~~~~~~Store ($ID_j, PID_j, key_j$) in the database      \\
&         $\underleftarrow {\{ID_j, PID_j, key_j, h(.)\}}$       &  \\
Store $\{ID_j, PID_j, key_j, h(.)\}$ in the memory & & \\
\hline
\end{tabular}}
\caption{drone registration of Nikooghadam et al.' scheme [6]} \label{fig:UPReg}
\end{figure}

\begin{figure}[!htbp]
\centering \scalebox{0.6}{
\hskip-3.03cm
\begin{tabular}{|l l l l l|}
\hline
~~~~~~~~~~~~\textbf{User $U_i$/Mobile device}&~~~~~~~~\textbf{Public Channel}~~~&~~~~~~~~~~~~~~~~\textbf{ControlServer($CS$)}&~~~~~~~~~~~\textbf{Public Channel}&~~~~~~~~~~~~~~~~~~~~ \textbf{Drone $V_j$}  \\
\hline
Input the identity $ID_i$ and password $PW_i$   & & & &  \\
Compute $ppw^\ast _i=h(h(PW_i||d_i)\oplus h(ID_i||d_i))$   & &   & &  \\
Compute $FID^\ast _i=h(ID_i||f_i)$   &  &  & &  \\
Compute $A^\ast _i=h(FID_i||ppw^\ast _i||f_i||K_i)\oplus h(ID_i||d_i))$   &  & & &  \\
Compute $B^\ast _i=h(A^\ast_i||FID^\ast _i)$ &  & & &  \\
If ($B^\ast _i \neq B_i$), reject the session &  & & &  \\
Else, select a timestamp $T_1$ &  & & &  \\
Select random number $z_i\in Z_p$ &  & & &  \\
Compute $A1_i=h(T_1||FID_i||K_i)$   & & & &  \\
&$\underrightarrow{\{T_1, z_iP, A1_i,FID_i, PID_j\}}$ & &   &   \\
& & Select timestamp $T_2$ & & \\
& & If ($|T_2-T_1| > \Delta T$), reject the session & &          \\
& & Else, retrieve ($ID_i, FID_i, K_i$) from the database & & \\
& & Compute $A1^\prime_i=h(T_1||FID_i||K_i$) & & \\
& & If ($A1^\prime_i\neq A1_i$), reject the session  & &      \\
& & Else, compute $K_{ij}=K_i\oplus key_j$ & & \\
& & Compute $A3_i=h(PID_j||key_j||ID_j||K_i$)& & \\
& &  &$\underrightarrow{\{A3_i, T_2, z_iP, PID_i, K_{ij}, FID_i\}}$   &   \\
& & & & Select timestamp $T_3$ \\
& & & & If ($|T_3-T_2|>\Delta T$), reject the session \\
& & & & Else, compute $K_i=K_{ij}\oplus key_j$ \\
& & & & Compute $A3_j=h(PID_j||key_j||ID_j||K_i)$ \\
& & & & If ($A3_j \neq A3_i$), reject the session \\
& & & & Else, select random number $g_j\in Z_p$ \\
& & & & Computer $sk_j,=h(ID_j||g_jz_iP||K_i|FID_i)$ \\
& & & & Compute $Auth_j=h(sk_j||FID_i||T_3||K_i)$ \\
& &&~~~~~~~$\underleftarrow{\{g_jP, T_3, Auth_j\}}$ &  \\
Select timestamp $T_4$  & & & &  \\
If ($|T_4-T_3|>\Delta T$), reject the session & &   & &  \\
Else, compute $sk_i=h(ID_j||z_ig_jP||K_i||FID_i$) &  &  & &  \\
Compute $Auth_i=h(sk_i||FID_i||T_3||K_i$) &  &  & &  \\
If ($Auth_i\neq Auth_j$), reject the session &  &  & &  \\
Else, authenticate $V_j$ &  &  & &  \\
Accept $sk_i(=sk_j)$ as the session key &  &  & &Accept $sk_j(=sk_i)$ as the session key  \\
\hline
\end{tabular}}
\caption{Login and authentication phase of Nikooghadam et al.' scheme [6]} \label{fig:UPReg}
\end{figure}

\subsection{Cryptanalysis of Nikooghadam et al.'s scheme}
In this section, we demonstrate that the scheme proposed by Nikooghadam et al. [6] suffers from user tracking and stolen verifier attacks.
\subsubsection{User tracking attack}
When user $u_i$ does the registration process, $CS$ Computes the identity parameter $FID_i = h(ID_i|| f_i)$, and This parameter is fixed during the protocol and does not change. So, when an attacker intercepts a user's login information $\{T_1 , z_iP , A1_i , FID_i , PID_j \}$, Afterwards can track the user's visit behavior with the help of the parameter $FID_i$.

\subsubsection{Stolen verifier attack - User impersonation}
Based on the assumption of the stolen verifier attack, the control server's database leaks, and stored information becomes available to the attacker; then, he attempts to impersonate the protocol parties. In Nikooghadam et al.'s scheme, the attacker obtains the parameters $T_1$ and $FID_i$ by intercepting data on the public channel and gets access to the value $K_i$ in the leaked database. As a result, the attacker can create parameter $A1_i=h(T_1||FID_i||K_i)$. When the control server checks whether $A1_i$ is equal to $A1^\prime_i$, not able to understand that this parameter is fraudulent. Therefore the attacker can take impersonate the User for the controller server.

\subsubsection{Stolen verifier attack - Server impersonation}
In a stolen verifier attack, the database of the control server is accessible to the attacker. Based on this assumption, by intercepting the public channel, the attacker obtains $PID_j$ and then access parameters $key_j, ID_j, K_i$ from the control server database. So the attacker can create parameter $A3_i=h(PID_j||key_j||ID_j|| K_i)$ and send it to the drone, whereas the drone can not distinguish the fake parameter $A3_i$ when verifying it. As a result, the attacker can impersonate the controller server for the drone.

\section{Concludion}
Providing a secure communication channel in the internet of drones has gained lots of attention. In this article, we reviewed the authentication protocol proposed by Nikooghadam et al. and demonstrated that it is prone to user tracking and stolen verifier attacks. In future, we plan to present a secure key agreement scheme for IoD that addresses the shortcomings of related works.

\end{document}